\documentclass[conference]{IEEEtran}
\IEEEoverridecommandlockouts
\usepackage{cite}
\usepackage{amsmath,amssymb,amsfonts}
\usepackage{algorithmic}
\usepackage{graphicx}
\usepackage{textcomp}
\usepackage{xcolor}
\def\BibTeX{{\rm B\kern-.05em{\sc i\kern-.025em b}\kern-.08em
    T\kern-.1667em\lower.7ex\hbox{E}\kern-.125emX}}
    
\usepackage{soul}
\usepackage{subfigure}
\usepackage{dblfloatfix}
\usepackage[margin=0.7in]{geometry}

\usepackage[framemethod=tikz]{mdframed}

\newcommand{\mbf}[1]{\mathbf{#1} } 

\newcommand{\I}{{\cal{I}}}
\newcommand{\X}{{\cal{X}}}
\renewcommand{\H}{{\cal{H}}}

\newcommand{\C}{{\cal{C}}}

\newcommand{\E}{{\mathbb{E}}}
\newcommand{\fig}[1]{Fig.~\ref{fig:#1}}
\newcommand{\eq}[1]{(\ref{eq:#1})}
\newcommand{\vect}[1]{\boldsymbol{#1}}

\newtheorem{theorem}{Theorem}

\begin{document}

\title{Capacity and Achievable Rates of Fading Few-mode MIMO IM/DD Optical Fiber Channels\\
\thanks{The authors are supported by Danish National Research Foundation Centre of Excellence SPOC,
DNRF123, and the Villum YI, OPTIC-AI, 29344.}
}

\author{
\IEEEauthorblockN{Metodi P. Yankov, Francesco Da Ros, Søren Forchhammer, and Lars Gr\"{u}ner-Nielsen}
\IEEEauthorblockA{\textit{Department of Photonics Engineering} \\
\textit{Technical University of Denmark}\\
2800 Kgs. Lyngby, Denmark \\
\{meya, fdro, sofo, larsgr\}@fotonik.dtu.dk} }

\maketitle

\begin{abstract}
The optical fiber multiple-input multiple-output (MIMO) channel with intensity modulation and direct detection (IM/DD) per spatial path is treated. The spatial dimensions represent the multiple modes employed for transmission and the cross-talk between them originates in the multiplexers and demultiplexers, which are polarization dependent and thus time-varying. The upper bounds from free-space IM/DD MIMO channels are adapted to the fiber case, and the constellation constrained capacity is constructively estimated using the Blahut-Arimoto algorithm. An autoencoder is then proposed to optimize a practical MIMO transmission in terms of pre-coder and detector assuming channel distribution knowledge at the transmitter. The pre-coders are shown to be robust to changes in the channel.   
\end{abstract}


\section{Introduction}
Few-mode transmission in optical fibers is typically considered in long-haul communication links in order to go beyond the so-called nonlinear Shannon limit of single-mode fibers and enhance the data rate by employing spatial multiplicity \cite{FiberMIMO, FiberMIMO_2}. On the other hand, for short-reach connections as the ones inside data centers (DCs), the problem is typically not capacity per fiber, but rather the cost of the transceivers, which have to employ multiple wavelengths to reach the target data rates. Such links typically operate with intensity-modulation and direct detection (IM/DD), with digital signal processing (DSP) functionalities kept at a minimum in order to reduce the cost w.r.t. the standard long-haul coherent transmission, and also to reduce latency which is critical for intra-DC communication. 

The abundance and low cost of few-mode fibers (FMFs) and the lifted requirement for frequency stabilization of the lasers make few-mode multiplexing an attractive alternative to few-wavelength multiplexing. The challenge in employing FMF transmission is  multiplexing and demultiplexing the transmission modes into and out of the same fiber. For that purpose, several types of mode multiplexers may be employed, e.g. air clad photonic lanterns \cite{GrunerOpex}, demonstrating 100G transmission over up to 2 km for 2-mode multiplexing. Due to the cost requirement, multiple input multiple output (MIMO) processing and synchronization of the modes at the receiver is typically prohibitive. Instead, the multiplexer (MUX) and demultiplexer (DEMUX) are required to produce negligible cross-talk (XT), which sets a harsh requirement on their fabrication process. 

Capacity and achievable rates for wireless IM/DD links have been studied\cite{LapidothCapacity}, more recently in \cite{Ndjiongue, ChaabanScalar}, including MIMO versions \cite{ChaabanMIMO, ChaabanMAC}, where the channel is assumed fixed and known at the transmitter and/or receiver. In general, the amplitude-constrained channel capacity study dates back to \cite{Smith}, where a linear system was considered. Achievable rate regions for the MIMO multiple access channel with IM/DD were derived in \cite{ChaabanMAC}. 

In this paper, the case of few-mode point-to-point optical fiber channel is considered with joint or independent mode processing at the receiver. The latter case falls into the category of MIMO broadcast channels with IM/DD, for which to our knowledge, no closed-form expressions exist for the channel capacity, especially in the case of fading. An upper bound on the capacity with perfect channel knowledge at the transmitter and receiver is adapted from the free-space case, and the Blahut-Arimoto algorithm (BAA) is used to estimate a tightly-approaching achievable rate. An autoencoder is then proposed to optimize a pre-coder and a detector for both cases of joint and independent processing at the receiver, under the constraint of only channel state distribution knowledge at the transmitter. 

\subsection{A note on notation}
Variables are denoted with capital letters, e.g. $X$, their distributions by $p_X(X)$, their support set by $\X$, with $\X_i$ being the $i-$th element, and their realizations by small letters, e.g. $x$ for scalars and $\mbf{x}$ for vectors, $x_i$ being the $i-$th vector element. Matrices are denoted $\mbf{H}$, to be distinguished from vector random variables $\vect{X}$.

\section{Channel model}
\begin{figure}
 \centering
 \includegraphics[width=0.95\linewidth]{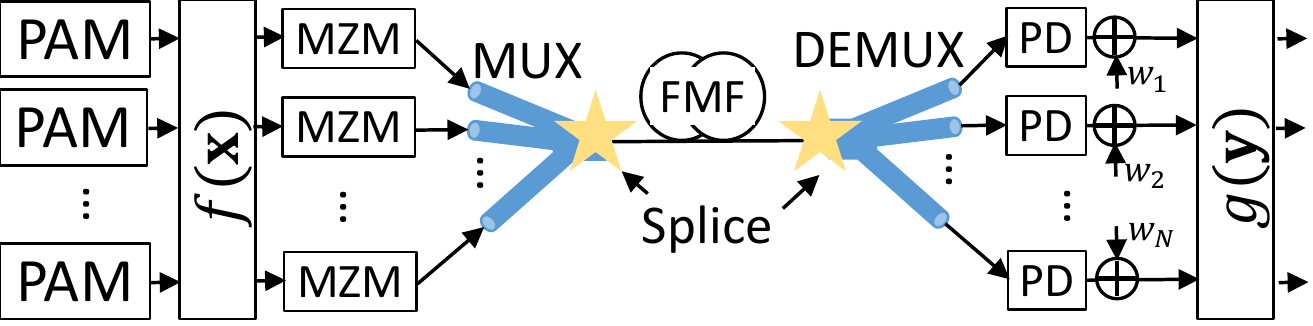}
 \caption{Considered MIMO IM/DD channel model of a few-mode fiber transmission.}
 \vspace{-0.5cm}
 \label{fig:channel}
\end{figure}

A transmission system with independent data on each spatial path of the FMF is considered. The data bits modulate the waveform using pulse amplitude modulation (PAM), where the modulation alphabet is $\X = {0, 1, ... M-1}$ of size $|\X|=M$. The symbols are independent in time and the signal vector is denoted $\mbf{x}=\left[x_1, x_2, ... x_N\right]^T$, where $N$ is the number of employed modes. The signal may then be pre-coded by a linear or nonlinear function $f$, $\mbf{v}=f(\mbf{x})$. A Mach-Zehnder modulator (MZM) is employed for modulating the intensity of optical carriers of wavelengths $\mbf{\lambda}=\{\lambda_1, \lambda_2, ... \lambda_N \}$ with the corresponding pre-coded symbols. The MZM is biased in quadrature, and its output intensity $s_i$ as a function of the modulating voltage $v_i$ can be modeled as
\begin{gather}
\label{eq:MZMCos}
 s_i = P_i \cos^2\left(\frac{\pi}{2}\frac{v_i}{V_{\pi}}\right),
\end{gather}
where $V_{\pi}$ is the half-wave voltage of the modulator, and $P_i$ is the optical power of the $i-$th carrier. The multiple fibers leading out of each MZM are combined in a mode MUX. An example of a MUX is the air clad photonic lantern made of single mode fibers, which are fused together and down tapered. The end of the taper is spliced to a few-mode fiber. After down tapering the surrounding air will act as the cladding. By use of single mode fibers with different core sizes mode selectivity is obtained \cite{GrunerOpex, GrunerOFC}. At the end, a DEMUX is used to demultiplex the modes into independent spatial paths. Each mode is detected by a photodetector (PD), which translates light intensity to electrical current. The signal at the receiver is denoted $\mbf{y}=\left[y_1, y_2, ... y_N\right]^T$. A block diagram of the system is given in \fig{channel}. Relatively short fiber distance is assumed so that dispersion, loss and distributed XT along the fiber span can be neglected \cite{GrunerOFC}. The channel is thus memoryless. Without loss of generality and for easier notation, the case of 2-mode transmission is assumed where the spatial modes $LP_{01}$ and $LP_{11}$ are excited for propagation. This case corresponds to the lanterns demonstrated in \cite{GrunerOpex}. 

Each of the MUX, DEMUX and splices can then be modeled by a MIMO matrix $\mbf{H}_{MUX}$, $\mbf{H}_{DEMUX}$ and $\mbf{H}_{SPL}$, respectively and has the form (for example for $\mbf{H}_{MUX}$)
\begin{equation}
\label{eq:basicMatrix}
 \mbf{H}_{MUX} = \begin{bmatrix} (1-xt_1) / \alpha_1  & xt_2 / \alpha_2 \\ xt_1 / \alpha_1  & (1-xt_2) / \alpha_2 \end{bmatrix} ,
\end{equation}
where $xt_1$ and $xt_2$ are the XTs from mode $LP_{01}$ to mode $LP_{11}$ ($LP_{11}$ to $LP_{01}$, respectively), and $\alpha_1$ and $\alpha_2$ are the losses of the corresponding modes in the component. The transfer matrices are subject to polarization-dependent XT. In this work, without loss of generality, this dependence is modeled by drawing the realization of each XT value in \eq{basicMatrix} for each time instant from a uniform distribution with a given mean $\E\left[XT\right]$ and a range of $6$ dB. The losses are assumed constant. Assuming the carriers are not coherent at the receiver (which is reasonable for cheap, independent lasers with significant phase noise and assuming large enough differential group delay, as well as proper wavelength selection for the carriers \cite{GrunerPTL}), the beating between modes can be minimized, and the channel from transmitter to receiver can be modeled as
\begin{gather}
\label{eq:MIMOChannel}
\mbf{y} = \mbf{H}_{DEMUX} \cdot \mbf{H}_{SPL}^2 \cdot \mbf{H}_{MUX} \cdot \mbf{s} +\mbf{w} = \mbf{H} \cdot \mbf{s}+\mbf{w} ,
\end{gather}
where $\mbf{s} = \left[s_1, s_2, ... s_N\right]^T$. The noise terms $\mbf{w}=\left[w_1, w_2, ... w_N\right]^T$ are independent, identically distributed samples from a Gaussian distribution with a variance $\sigma_w^2$. The noise is assumed dominated by the receiver thermal and quantization noise. At the receiver, independent or joint detection of the signal $\mbf{y}$ may be performed with the function $g(\mbf{y})$.

The MZM transfer function imposes an amplitude constraint on the signal $0 \le v_i \le V_{\pi}$, since outside of these boundaries the cosine transfer function becomes non-bijective. Equivalently, the channel in \eq{MIMOChannel} can be considered the standard MIMO IM/DD channel \cite{ChaabanMIMO, ChaabanMAC} with input $\mbf{s}$ and amplitude constraint $0 \le s_i \le P_i$.

In the case of fiber communications and for IM/DD links operating at short reach, there is no practical constraint (within reason) associated with the average power, unlike for example free-space optical wireless links, where safety and interference may be factors.

The signal to noise ratio (SNR) is defined as $SNR=10\log_{10}( \frac{\sum_iP_i}{2N\cdot\sigma_w})$, motivated by the fact that $\E\left[|Y_i-S_i|\right]=\sigma_w$ and that for uniformly distributed $s_i$ we have $\E\left[|S_i|\right]=P_i/2$. A total power constraint may be imposed in the form of $\sum_i P_i = P_{tot}$ so that the SNR can be swept. 

\section{Capacity and capacity achieving distributions}
\label{sec:capacity}
Assuming full-rank, positive $\mbf{H}$ with well-behaved distribution, the ergodic capacity $\C$ of the MIMO fading channel with perfect channel knowledge at the transmitter and receiver is defined as
\begin{gather}
\label{eq:capDef}
\C=\max_{p_{\vect{X}}(\vect{X}), f({\vect{X}}), P_{\{1\dots N\}}}\E_H\left[\I(\vect{Y}; \vect{X}|\mbf{H}, p_{\vect{X}}(\vect{X}))\right], 
\end{gather}
where the mutual information (MI) $\I(\cdot; \cdot)$ between the input and output is maximized over the input distribution $p_{\vect{X}}(\vect{X})$, the pre-coder $f({\vect{X}})$ and the power allocation. Assuming that the optimal pre-coder is applied, we have
\begin{gather}
\label{eq:capDefEquiv}
\C=\max_{p_{\vect{S}}(\vect{S}), P_{\{1\dots N\}}}\E_H\left[\I(\vect{Y}; \vect{S}|\mbf{H}, p_{\vect{S}}(\vect{S}))\right],
\end{gather}
which is the ergodic capacity of the channel in \eq{MIMOChannel} with ideal knowledge at the receiver. The latter can be upper bounded by splitting the channel into parallel IM/DD channels by using the QR decomposition of $\mbf{H=QR}$, where $\mbf{Q}$ is unitary and $\mbf{R}$ is upper triangular. The receiver then pre-processes $\mbf{y}$ to $\hat{\mbf{y}}=\mbf{Q}^T\mbf{y}=\mbf{R}\mbf{s}+ \mbf{w}$, the interference from previously decoded channels is canceled on sub-sequent channels and the MIMO channel is equivalent to the set of parallel channels $\hat{y}_i=\mbf{R}_{i,i}s_i+w_i$, where $\mbf{R}_{i,i}$ is the element on the $i-$th row and $i-$th column on $\mbf{R}$ \cite{ChaabanMIMO}.  The capacity may then be upper-bounded by
\begin{gather}
\label{eq:capUpBound}
 \C \le \max_{P_{\{1\dots N\}}}\E_H\left[ \sum_i \bar{\C}_i(\mbf{R}_{i,i}, P_i, \sigma)\right],
\end{gather}
where $\bar{\C}_i(\mbf{R}_{i,i}, P_i, \sigma)$ is the upper bound (UB) on the scalar IM/DD channel with power $P_i$ and a gain coefficient $\mbf{R}_{i,i}$. The maximization over the power allocation can be done by a grid search for reasonable number of spatial modes with the above mentioned constraint of $\sum_i P_i = P_{tot}$. Two upper bounds are considered, UB1 (obtained from \cite[Theorem 1]{ChaabanScalar}) and UB2 (obtaiend from \cite[Eq. \(19\)]{LapidothCapacity}), where UB1 is tight at high SNR, and UB2 is superior at low SNR. The limitation of these UBs is that they do not provide the pre-coder function, since they are derived in the already pre-coded space. 

On the other hand, practically achievable capacity estimates can be derived numerically using the classical BAA. If the optimal distribution for the channel in \eq{MIMOChannel} is defined as $p_{\vect{S}}^{opt}(\vect{S})=\sum_{k=1:|{\cal{S}}|} a_k \delta(S-{\cal{S}}_k)$, where $a_k$ are the weights of the $k-$th mass point ${\cal{S}}_k$ and $\delta(\cdot)$ is the Dirac delta function, the following can be proven. 
\begin{theorem}{For a given power allocation $P_{\{1 \dots N\}}$, an initialization of the BAA with a discrete uniform distribution of $M\rightarrow \infty$ on the constrained hypercube produces $p_{\vect{S}}^{opt}(\vect{S})$ for which the MI $\E_H\left[\I(\vect{Y}; \vect{S}|\mbf{H}, p_{\vect{S}}^{opt}(\vect{S}))\right]$ is the capacity of the channel in  \eq{MIMOChannel}, and a lower bound on $\C$. }
\begin{IEEEproof}
The following observations can be made:
\begin{enumerate}
 \item The capacity-achieving distribution of channel \eq{MIMOChannel} is discrete and unique \cite{ChanPMFMeasure}\cite[Chapter 11]{CapacityBook}, and the MI is concave in $p_{\vect{S}}(\vect{S})$ \cite{Faycal, LapidothCapacity}.
 \item For discrete, memoryless channels the BAA achieves improvement on every iteration \cite{Cover} until convergence (i.e., it always achieves a local optimum).
 \item The BAA is allowed to approach zero probability to individual mass points, effectively removing points from the support set of $\vect{S}$.
 \item Since the power constraint is ensured by the MZM transfer function \eq{MZMCos} and the initialization of the mass points, the BAA is \textit{unconstrained} (except for $\sum_i a_i = 1$).
\end{enumerate}
From Observations 3) and 4), for $M \rightarrow \infty$ the BAA can 'reach' any discrete support set within the hypercube, making the corresponding local optimum from Observation 2) the global optimum due to the concavity in Observation 1), proving the first part of the Theorem. The MI achieved with the BAA is an achievable rate on the original channel with the trivial pre-coder $\mbf{V}=f(X)=\mbf{X}$ and support set ${\cal{V}}$ obtained by inverting \eq{MZMCos}, making it by definition a lower bound on the capacity in \eq{capDef}, proving the second part of the Theorem
\end{IEEEproof}
\end{theorem}
Theorem 1 has been used implicitly to estimate the capacity of the scalar IM/DD channel \cite{CapacityBook}, and is here formalized and proven. Furthermore, we point out that initialization with a distribution of any size $M$ and any selection of mass points ${\cal{S}}_k$ also provides an achievable rate, albeit potentially lower. A particularly convenient and relevant initialization is for $\vect{V}$ to be uniformly distributed on the hypercube confined by $\left[0; V_{\pi}\right]^N$ and $\vect{S}$ obtained by applying \eq{MZMCos}, which allows for the potential shaping gain to be estimated w.r.t. classical equidistant PAM$^N$ schemes (as shown below). 


Some capacity estimates are provided in \fig{cap} for initializations as described above of different size, together with the two considered UBs. The expected channel parameters in this case are given in Table~\ref{tbl:params}, and correspond to the measurements done in \cite{GrunerOFC}. For the rest of the paper, the total power constraint is arbitrarily chosen to be $\sum_i P_i=3$ dBm. In certain SNR regions of interest (between $\approx$~15 dB and $\approx$~20 dB), the BAA constrained to 32PAM per dimension closes the gap to the upper bound, effectively achieving the channel capacity with the trivial pre-coder described above and ideal channel knowledge at the receiver. The shaping gain for this constellation size is limited to $\approx$~0.5 dB w.r.t. the MI achieved with uniformly distributed, equidistant points. As shall be seen in the Results section, when the constellation size remains fixed and the SNR is increased above a certain modulation-size dependent threshold, pre-coding plays a more important role and will influence the tightness of the bound provided by the rate, achievable with the BAA. 

\begin{table}[!t]
 \caption{Expected parameters for the MIMO components.}
 \label{tbl:params}
 \centering
 \begin{tabular}{r||c|c|c|c}
  & $\E\left[XT_1\right]$, dB & $\E\left[XT_2\right]$, dB & $\alpha_1$, dB & $\alpha_2$, dB \\
 \hline
  MUX & -18 & -15 & 0.7 & 1.4 \\  
  DEMUX & -11 & -11 & 1.5 & 3 \\
  SPL & -25 & -25 & 0.04 & 0.04 \\
 \hline
 \hline
 \end{tabular}
 \vspace{-0.5cm}
\end{table}

\begin{figure}[!t]
 \centering
 \includegraphics[width=0.95\linewidth]{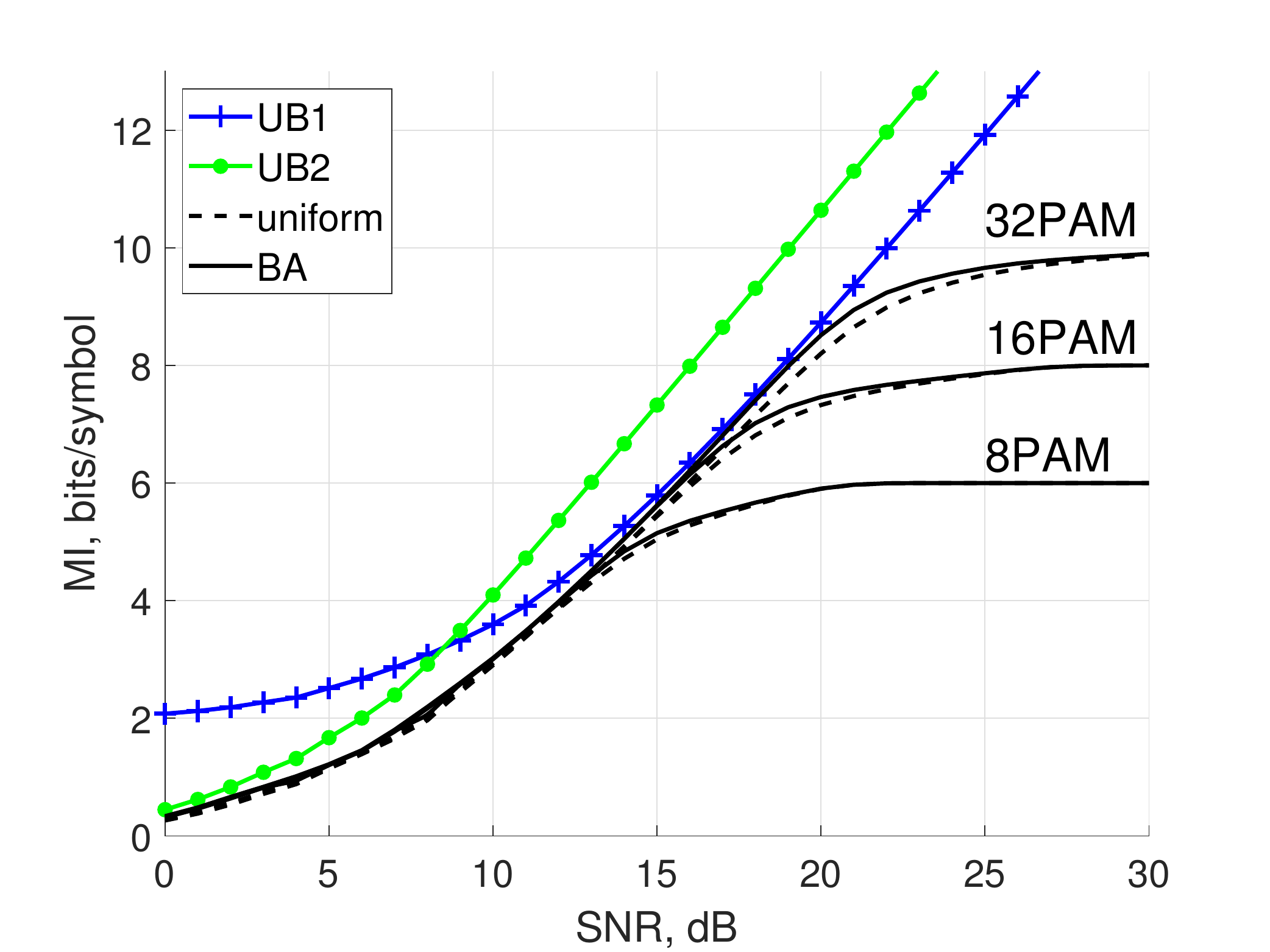}
 \caption{Capacity results for the parameters given in Table~\ref{tbl:params}. }
 \vspace{-0.5cm}
 \label{fig:cap}
\end{figure}

\begin{figure}[!t]
 \centering
 \includegraphics[width=0.95\linewidth]{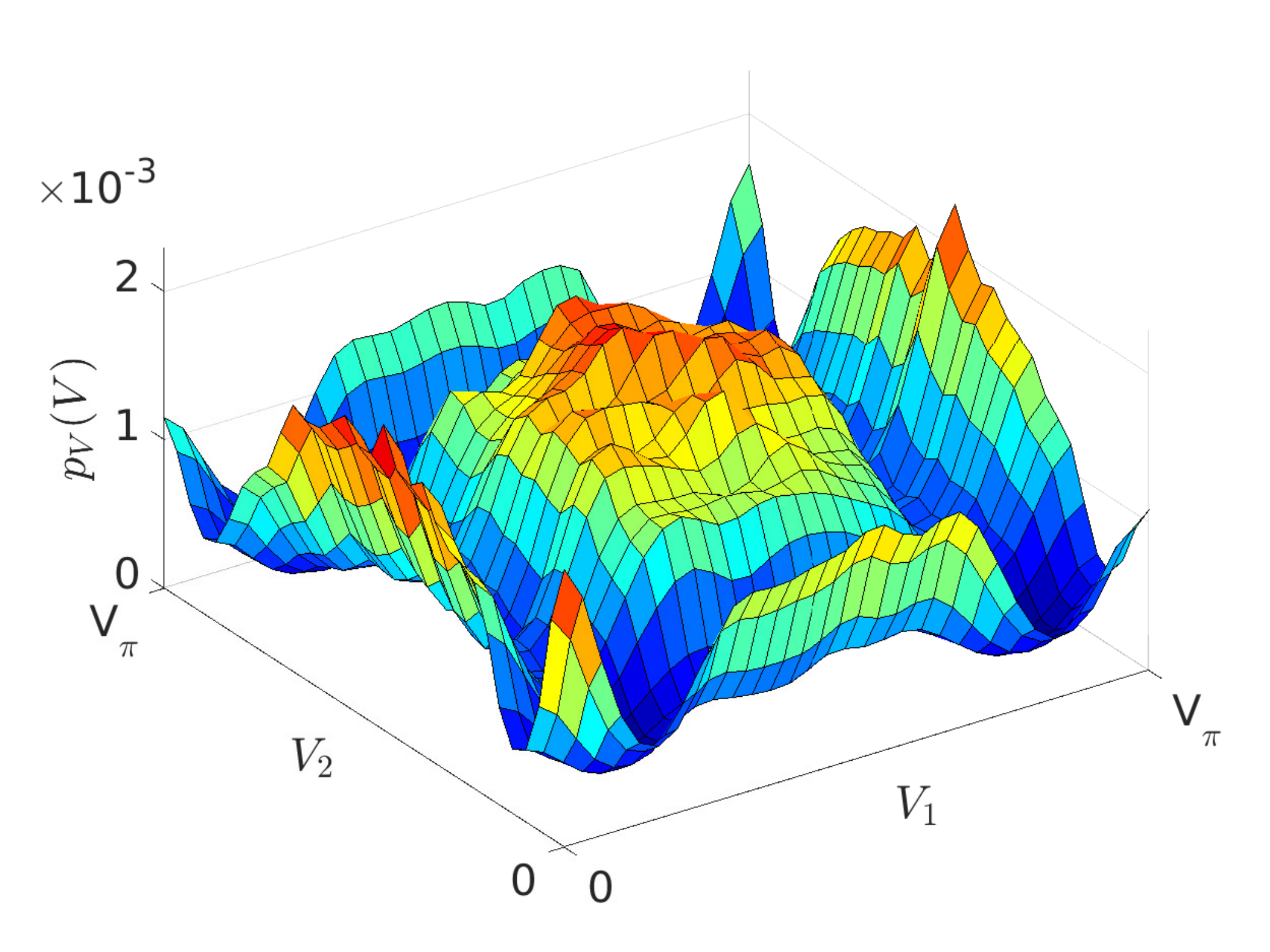}
 \caption{An example optimal distribution $p_{\vect{V}}(\vect{V})$ for $M=32$ and $SNR=15$ dB. }
 \vspace{-0.5cm}
 \label{fig:BAPMF}
\end{figure}

An example of the capacity achieving distribution for $M=32$ and $SNR=15$ dB is given in \fig{BAPMF}. We see a slight asymmetry in the distribution corresponding to the asymmetric XT. The optimal powers found by grid search (in this case a line search, since $N=2$ and $P_1+P_2=3$ dBm) were found to be $P_1=0.41$ dBm and $P_2=-0.46$ dBm.

\section{Autoencoder based communication}
As mentioned in the introduction, full MIMO processing may be too complex for the target communication system because of both hardware (in terms of synchronizing the receivers for each mode) and software (in terms of processing for channel estimation and multi-dimensional detection) requirements. However, partial channel knowledge may still be available in terms of the distribution of $\mbf{H}$ at the transmitter and receiver, which can be obtained e.g. from measurements statistics. For such cases, an auxiliary channel likelihood function must be employed for detection and MI estimation, which is subject to optimization. The likelihood function needs to be robust to the distribution of $\mbf{H}$. At the same time, the knowledge of $\mbf{H}$ may allow to design a pre-coder which at least partially mitigates the mode interference. Both of these optimizations can be efficiently approached with an autoencoder (AE) \cite{AE1}. The AE employs a neural network (NN) for pre-coding, another NN for detection, and wraps them around the channel model at hand. The most closely related to this work AE application is for the Rayleigh fading MIMO channel \cite{AERayl}, where the channel is linear. In this paper, both the MZM and the photodetection induce nonlinearities. Furthermore, in this paper, the AE is employed to also optimize the power allocation (instead of the grid-search performed in the previous Section). A basic schematic of the AE is given in \fig{AE}, and the following optimization scenarios are considered for the pre-coder:

\begin{figure}
 \centering
 \includegraphics[width=0.95\linewidth]{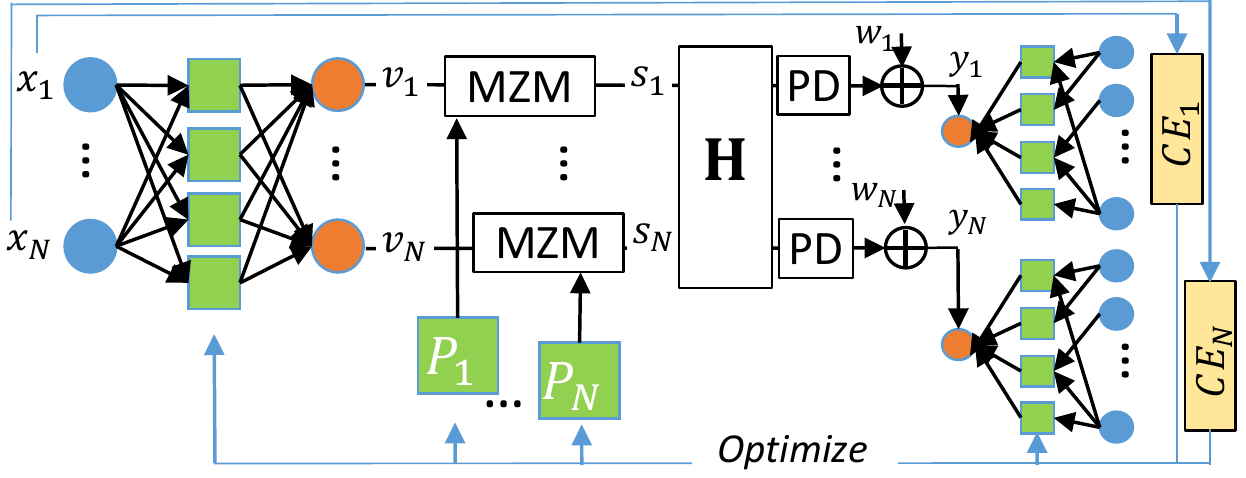}
 \caption{Autoencoder model for optimization. Showcased with independent detector NN per layer.}
 \vspace{-0.5cm}
 \label{fig:AE}
\end{figure}

\begin{description}
 \item[\textbf{Prec1}] Power allocation only. The pre-coder is $\mbf{v}=\frac{V_{\pi}}{M-1}\mbf{x}$. 
 \item[\textbf{Prec2}] Power allocation + linear pre-coder $\mbf{v}=f(\mbf{x})=\mbf{A}\mbf{x}$, where $A$ is an NxN matrix.
 \item[\textbf{Prec3}] Power allocation + nonlinear pre-coder, where $f(\mbf{x})$ is a multi-layer NN.
\end{description}

\noindent 
and for the detector:
\begin{description}
 \item[\textbf{Det1}] Each mode is detected by an independent NN.
 \item[\textbf{Det2}] All modes are detected jointly by a common NN.
 \end{description}
 
The outputs of the detector NNs are auxiliary distributions $q(\mbf{x}|\mbf{y})$ for the conditional distribution $p(\mbf{x}|\mbf{y})$, and the detectors thus provide an achievable rate \cite{Arnold}. \textbf{Det1} represents a more practical scenario. In order to simplify it even further, its performance is reported also for the case of a standard Gaussian auxiliary channel receiver, for which the auxiliary distribution is $q(\X_i|y_k) \propto p_{X_i}(\X_i) \cdot \exp \left[ -0.5/\hat{\sigma}_{w_i} \left( y_k-\mu_i\right)^2 \right]$. The parameters $\left(\mu_i, \hat{\sigma}_{w_i} \right)$ of the auxiliary Gaussian function are estimated using training data \cite{YankovJLT}. The scenario with \textbf{Prec1} and independent Gaussian detection per mode can be thought of as the current state of the art in few-mode, short reach IM/DD fiber links, and is our point of performance reference. In the case of \textbf{Prec2} and \textbf{Prec3} and Gaussian detection, a NN is still employed at the receiver but only in order to facilitate training of the pre-coder using the cross-entropy (CE) cost function. 

In the case of \textbf{Det1}, the AE is trained using the cost function $Cost = \max_{i=1 \dots N} CE_i(X_i, g(Y_i))$, where $CE_i(X_i|Y_i) = \H(X_i|Y_i)$ is the CE, which in this case is equivalent to the conditional entropy $\H(X|Y)$. Since the MI $\I(X;Y) = \H(X)-\H(X|Y)$, this cost function targets equal performance on all modes, which is the desired mode of operation in simple, robust systems which do not allow rate adaptivity. The case of \textbf{Det2} is considered as a reference performance and for comparison to the bounds and rates of Section~\ref{sec:capacity}. In this case, the cost function is $Cost = CE(\vect{X}, g(\vect{Y}))$. For all detectors and pre-coders, the NN topologies were optimized by increasing their depth and width by factors of 2 until convergence of the cost function. The resulting topologies for $N=2$ are given in Table~\ref{tbl:topologies}. The last layer of the detectors is a classifier, and thus has output size of $M$ and $M^2$ for the cases of independent and joint processing, respectively. All NNs employ the ReLU activation function and all NNs are biased. The Adam optimizer was employed for stochastic gradient descent optimization of all parameters. The batch size was 200, and a total of 1.000.000 symbols are used for training and independent 100.000 for testing. The channel is drawn from a uniform distribution as described above for every sample during training and testing. The result is an AE robust to the desired channel distribution. Observe, no explicit constraint is imposed on the encoder part. The AE is naturally forced to operate within the boundaries of the cosine transfer function of the MZM \eq{MZMCos}.

\begin{table}[!t]
 \caption{Optimized topologies in terms of nodes per layer of the AE components for $N=2$. All NNs are biased.}
 \label{tbl:topologies}
 \centering
 \begin{tabular}{r||c|c}
  & 8PAM & 16PAM \\
 \hline
  \textbf{Prec2} & [2x2] & [2x2] \\  
  \textbf{Prec3} & [2x16x16x2] & [2x16x16x2] \\ 
  \textbf{Det1} & [1x64x64x8] & [1x256x256x16] \\ 
  \textbf{Det2} & [2x128x128x64] & [2x512x512x256] \\ 
 \hline
 \hline
 \end{tabular}
 \vspace{-0.5cm}
\end{table}

\section{Results}
\begin{figure}
 \centering
 \includegraphics[width=0.95\linewidth]{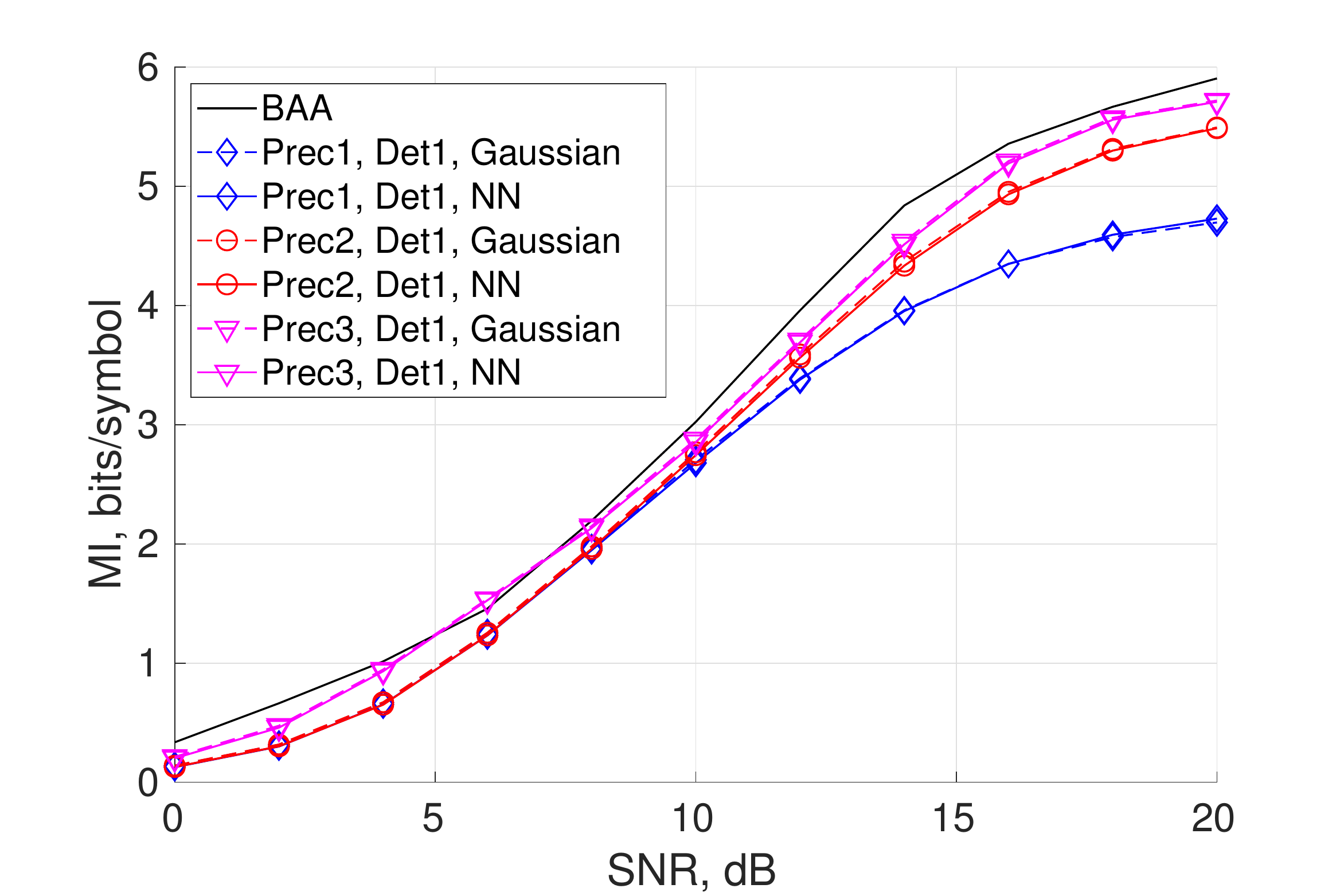}
 \vspace{-0.2cm}
 \caption{AE performance for the channel with expected parameters from Table~\ref{tbl:params} and uniform distribution, at $M=8$. The rate is estimated both with the NN detector and with a Gaussian receiver, both with independent processing per mode (\textbf{Det1}).}
 \vspace{-0.5cm}
 \label{fig:indepProcessing}
\end{figure}

\begin{figure}
 \centering
 \includegraphics[width=0.95\linewidth]{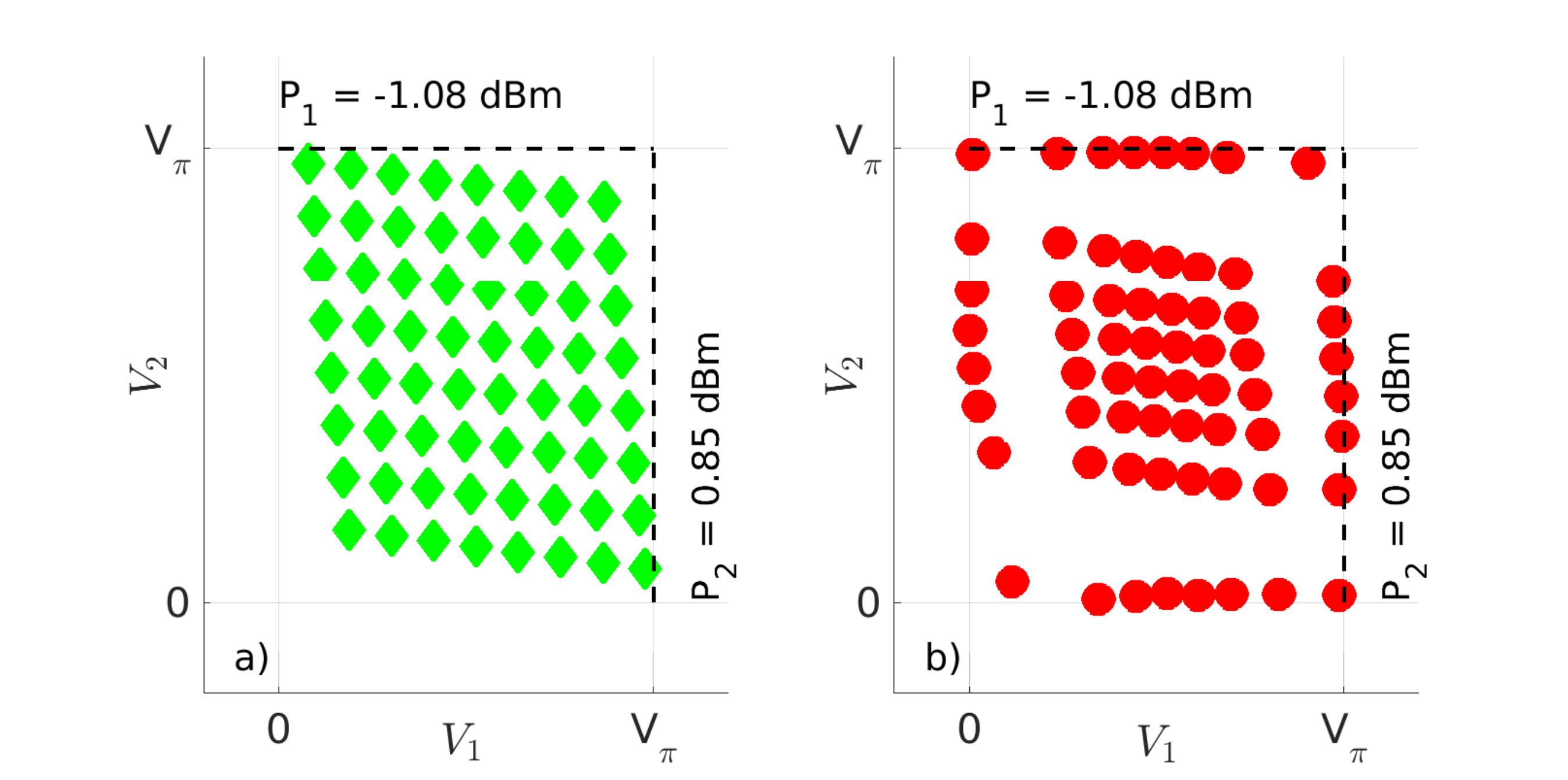}
 \caption{Precoded constellations for the channel with expected parameters from Table~\ref{tbl:params} and uniform distribution, at $M=8$ and $SNR=15$ dB. \textbf{a): Prec2} (linear); \textbf{b) Prec3} (nonlinear).}
 \vspace{-0.5cm}
 \label{fig:constellations}
\end{figure}

The practically-relevant system with \textbf{Det1} is first treated where independent mode processing is employed at the receiver in \fig{indepProcessing} for $M=8$. The reported rate corresponds to $N \cdot \min_i \E_H \left[ \I(Y_i;X_i | p_{\mbf{H}}(\mbf{H})) \right]$, i.e., only the \textit{distribution} of $\mbf{H}$ is known, and the performance of the worst mode is considered. The standard Gaussian receiver without additional pre-coding is limited in rate to $\approx 4.8$ bits/symbol at $SNR=20$ dB. The rate is substantially improved by the AEs for $SNR>10$ dB by up to $1$ bits/symbol at the high end. Alternatively, the SNR gain is $\approx 5$ dB at the maximum rate for the conventional system. The gain of nonlinear pre-coding is up to $\approx 0.25$ bits/symbol w.r.t. linear pre-coding, both at low and high SNR. As expected from the Gaussian statistics of the noise, there is no extra benefit from nonlinear detection w.r.t. standard Gaussian receiver. The BAA rate is also shown for comparison. In contrast, its rate is $\E_H\left[ \I(\vect{Y};\vect{X}| \mbf{H})\right]$. We see that nonlinear pre-coding operates near the BAA rate, demonstrating that the penalty from independent processing and lack of ideal knowledge of $\mbf{H}$ at the receiver is nearly recovered by simple, time invariant pre-coding. The pre-coded constellations are shown in \fig{constellations} for $M=8$ and $SNR=15$ dB. The linear pre-coder can only skew the constellation, whereas the nonlinear one can shape the constellation so that the distribution of energy is similar to that of \fig{BAPMF}. The optimal distribution obtains a shape, which takes advantage of the full-range of the modulation of $\left[0; V_\pi \right]$, while at the same time concentrates the mass around the center of the modulation support $\left(V_\pi/2; V_\pi/2 \right)$ in order to increase the robustness to the Gaussian noise. Similarly to the linear case, the nonlinear edges and skewing are attributed to the asymmetric expectation of the cross-talk. For reference, the optimal powers found by the AE are also indicated in \fig{constellations} for each case. The powers are now different from the ones obtained for the BAA because the target is equalizing the performance per mode, instead of the total sum-rate.

\begin{figure}
 \centering
 \includegraphics[width=0.95\linewidth]{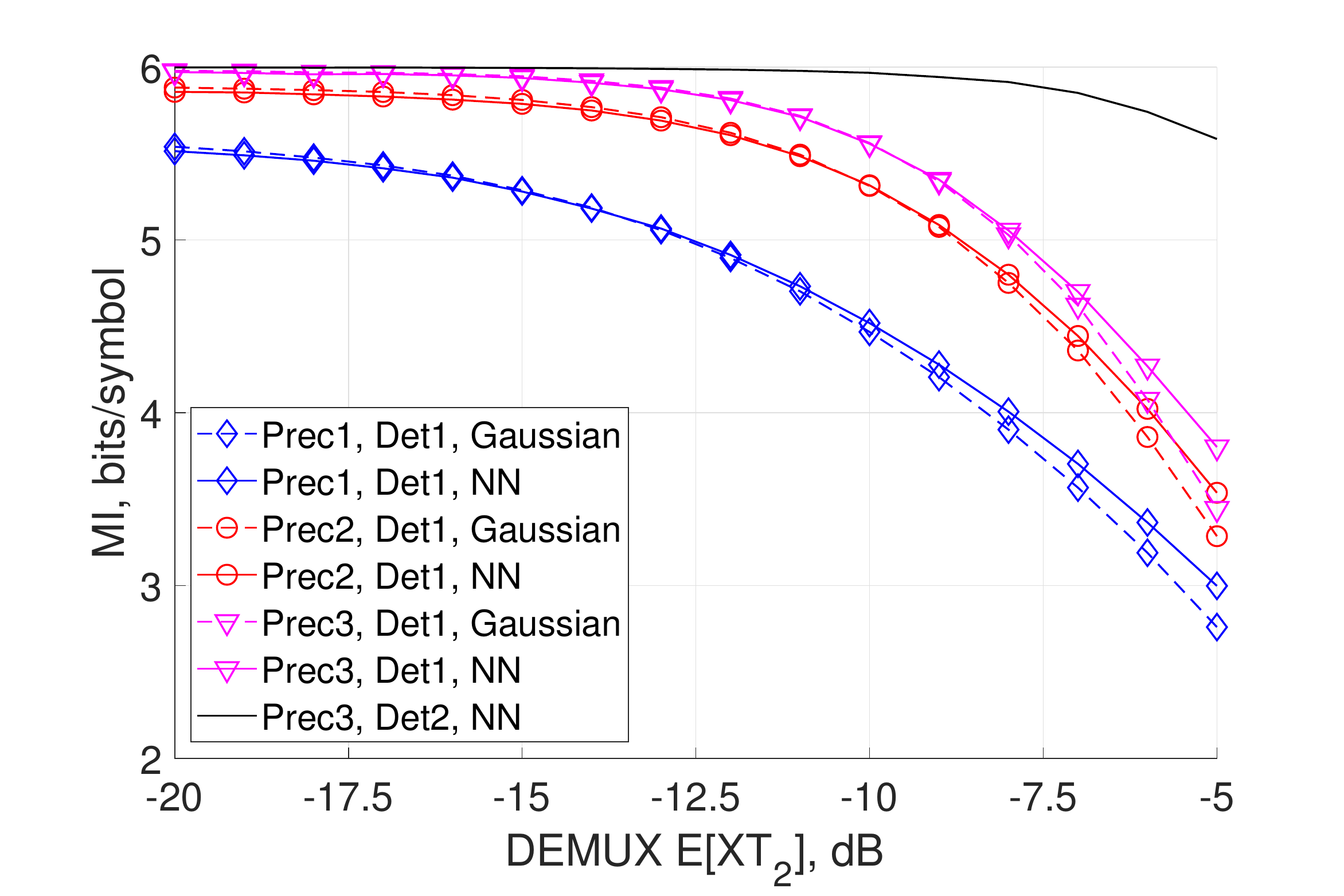}
 \vspace{-0.2cm}
 \caption{AE performance for the channel with expected parameters from Table~\ref{tbl:params} and uniform distribution, except for variable $XT_2$ of the DEMUX, at $M=8$, $SNR=20$ dB. The rate is estimated both with the NN detector and with a Gaussian receiver, both with independent processing per mode (\textbf{Det1}).}
 \vspace{-0.5cm}
 \label{fig:XT_study}
\end{figure}

The benefits of the AE w.r.t. the standard uncoded approach can also be exemplified in terms of XT tolerance. To that end, without loss of generality the expected $XT_2$ of the DEMUX is varied between -20 and -5 dB, and the rest of the parameters remain the same. The results at the $SNR=20$ dB are given in \fig{XT_study}. In order to achieve an example target rate of 5.094, which corresponds to a code rate of 0.849 of a typical passive optical network forward error correction low-density parity check code \cite{ZTE_FEC}, the standard system requires that the XT is reduced to $\approx-13$ dB. The proposed pre-coders achieve the target rate with XT tolerances of $4$ dB and $5$ dB for the linear and nonlinear pre-coder, respectively. This tolerance is directly translated to loosened fabrication requirements for the MUX and DEMUX. At the extreme end of the XT, the nonlinear detectors also exhibit gain w.r.t. the Gaussian receiver, potentially due to their capability to achieve an asymmetric likelihood function. 

\begin{figure}
 \centering
 \includegraphics[width=0.95\linewidth]{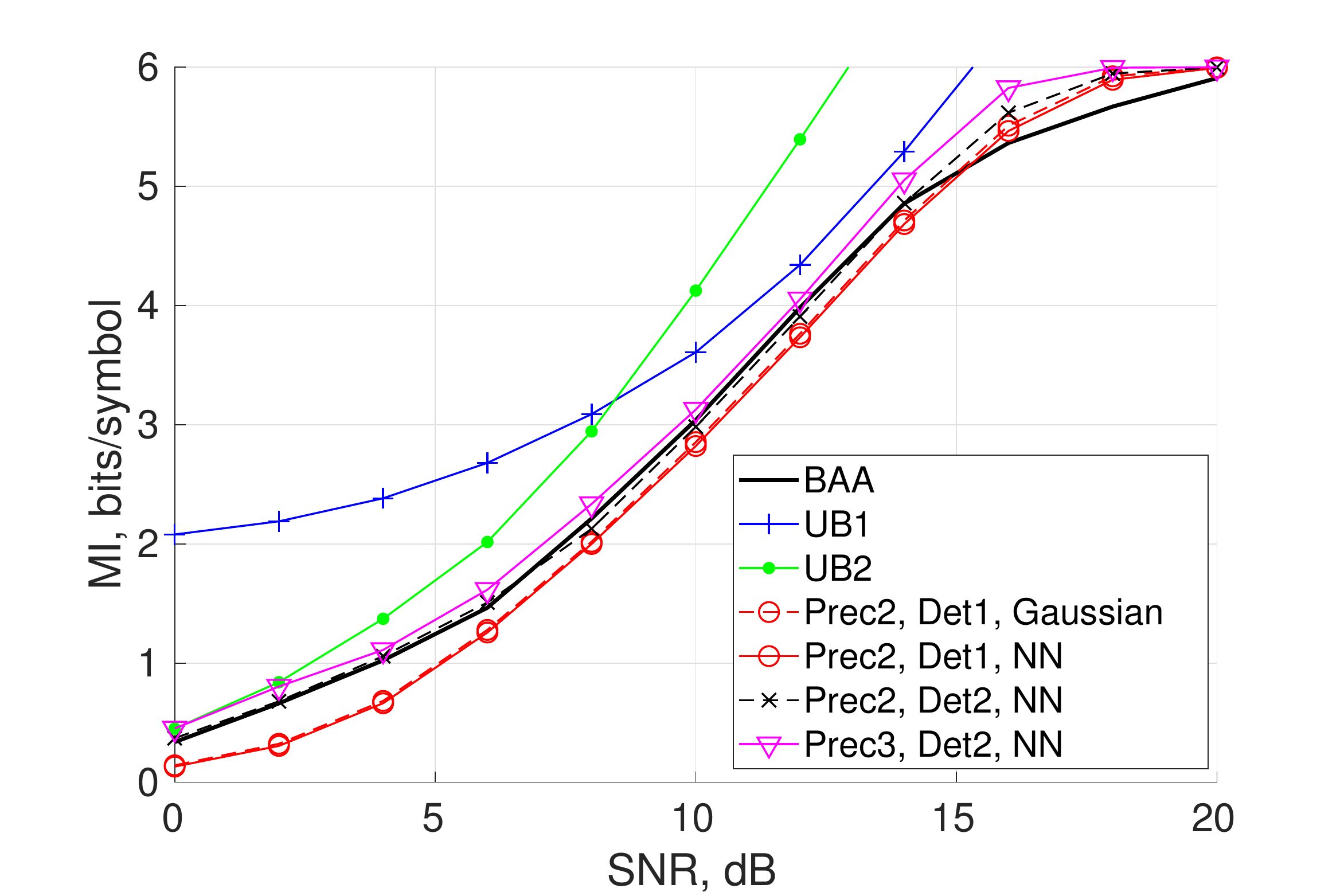}
 \caption{AE performance for the channel with expected parameters from Table~\ref{tbl:params} with 0 uncertainty, at $M=8$.}
 \vspace{-0.5cm}
 \label{fig:certainChannel}
\end{figure}

To further study the effect of the pre-coder, the rate of the AE is estimated and compared to that of the BAA for the case of 0 uncertainty in the channel in \fig{certainChannel}. This case still has relevance in practice if e.g. polarization controllers are employed at the transmitter and receiver to stabilize the polarization and remove the XT drift. The AE requirements for robustness to channel uncertainty are now relaxed, which means that it can find a better encoder targeted at the known channel. No such benefit can be exploited by the BAA assignment of probabilities since as mentioned above, it already operates in the pre-coded space. At the moderate SNR, the BAA is near-capacity achieving, as is the AE. However, at higher SNR and at rates above $\approx 5$ bits/symbol, the impact of the mode interference becomes dominant to that of the noise, and pre-coding becomes beneficial. The simplest AE with linear pre-coding and independent processing at the receiver already outperforms the BAA. Observe, the latter has access to the channel at the receiver. The full AE with nonlinear pre-coding and joint mode detector outperforms the BAA on the entire SNR range and closes the gap to capacity further. 

\section{Discussion}
The BAA optimization can be thought of as probabilistic shaping, while the pre-coding can be thought of as geometric shaping. An obvious direction for future research is to explore joint optimization of both, since as it was seen above, each has its benefits. 

In this work, uniform distribution of the cross-talk was arbitrarily chosen. The described methods directly support other types of distributions, e.g. Gaussian and or empirical, depending on the fit to the measured polarization drift.

The case of training independent detectors per mode with the target of maximizing the sum-rate instead of equalizing the rate per mode is well-within the realm of capabilities of the AE. It is relevant for cases where rate-adaptivity is permitted per mode. It was omitted here for space considerations.  

In this paper, $N=2$ in order to present some practically relevant and motivated results, directly applicable to the system of \cite{GrunerOFC}. Higher-order mode multiplexing with $N>2$ is well-supported by the presented concepts. 

As mentioned in the introduction, the capacity of the MIMO IM/DD multiple access channel has been studied before \cite{ChaabanMAC}. The channel presented in this work falls instead in the category of MIMO broadcast channels, since the receivers are independent. Tapping into the general broadcast MIMO channel capacity results, e.g. from \cite{MIMOBC} in order to improve the estimates and develop novel pre-coding techniques for the case of IM/DD transmission is of interest for future work. 

The air clad lanterns were only used as an example in order to provide practical cross-talk numbers. The bounds and AE method are general to other types of mode multiplexers, e.g. integrated ones \cite{integratedMUX}. 

\section{Conclusion}
The capacity of the optical FMF MIMO channel was lower-bounded by means of an unconstrained BAA, and tightness was demonstrated to previous upper bounds, resulting in accurate capacity estimates. The lower bounds are shown to be achievable by means of trivial pre-coders. Non-trivial pre-coders and detectors were then designed based on an AE. The AE with independent mode processing at the receiver was shown to approach the BAA lower bound and at the same time to be robust against channel uncertainties and drifts. For a fixed channel, the pre-coders were shown to be superior to the BAA and further close the gap to capacity for constellations of practical size. Furthermore, up to $5$ dB of cross-talk tolerance was reported by the AE w.r.t. standard techniques, which will enable cheaper multiplexers and demultiplexers to be employed for FMF transmission alternatives to standard wavelength division multiplexing of short reach optical links. 



\bibliographystyle{IEEEtran}
\bibliography{references}


\end{document}